\documentclass[preprint,journal]{vgtc}            

\onlineid{1147}

\preprinttext{To appear in IEEE Transactions on Visualization and Computer Graphics.}

\vgtccategory{Research}

\vgtcpapertype{model, evaluation}

\title{A Framework for Multimodal Medical Image Interaction}

\author{%
  \authororcid{Laura Schütz*}{0000-0002-5534-3903}, 
  \authororcid{Sasan Matinfar*}{0009-0003-6905-2001}, 
  \authororcid{Gideon Schafroth}{0009-0001-9651-4924}, 
  Navid Navab,
  \authororcid{Merle Fairhurst}{0000-0001-6540-5891},
  \\ \authororcid{Arthur Wagner, MD}{0000-0001-9947-2240},
  \authororcid{Benedikt Wiestler, MD}{0000-0002-2963-7772},
  \authororcid{Ulrich Eck}{0000-0002-5322-4724}, and \authororcid{Nassir Navab}{0000-0002-6032-5611}
}

\authorfooter{
  \item
  	Laura Schütz is with the Technical University of Munich (TUM).
  	\\E-mail: laura.schuetz@tum.de
  \item
  	Sasan Matinfar is with TUM.
  	E-mail: sasan.matinfar@tum.de.
   \item
  	Gideon Schafroth is with TUM.
  	E-mail: gideon.schafroth@tum.de.
   \item 
        Navid Navab is with Concordia University.
  	E-mail: navid.nav@gmail.com.
   \item 
        Merle Fairhurst is with TU Dresden.
  	E-mail: merle.fairhurst@tu-dresden.de.
   \item 
        Arthur Wagner is with TUM.
  	E-mail: arthur.wagner@tum.de.
   \item
  	Benedikt Wiestler is with TUM.
  	E-mail: b.wiestler@tum.de.
   \item
  	Ulrich Eck is with TUM.
  	E-mail: ulrich.eck@tum.de.
   \item
  	Nassir Navab is with TUM.
  	E-mail: nassir.navab@tum.de.
   \\ \text{*} contributed equally to this work.
}

\abstract{%
Medical doctors rely on images of the human anatomy, such as magnetic resonance imaging (MRI), to localize regions of interest in the patient during diagnosis and treatment. Despite advances in medical imaging technology, the information conveyance remains unimodal. This visual representation fails to capture the complexity of the real, multisensory interaction with human tissue. However, perceiving multimodal information about the patient's anatomy and disease in real-time is critical for the success of medical procedures and patient outcome. We introduce a Multimodal Medical Image Interaction (MMII) framework to allow medical experts a dynamic, audiovisual interaction with human tissue in three-dimensional space. In a virtual reality environment, the user receives physically informed audiovisual feedback to improve the spatial perception of anatomical structures. MMII uses a model-based sonification approach to generate sounds derived from the geometry and physical properties of tissue, thereby eliminating the need for hand-crafted sound design. Two user studies involving 34 general and nine clinical experts were conducted to evaluate the proposed interaction framework's learnability, usability, and accuracy. Our results showed excellent learnability of audiovisual correspondence as the rate of correct associations significantly improved ($p<0.001$) over the course of the study. MMII resulted in superior brain tumor localization accuracy ($p<0.05$) compared to conventional medical image interaction. Our findings substantiate the potential of this novel framework to enhance interaction with medical images, for example, during surgical procedures where immediate and precise {feedback} is needed.
}

\keywords{Multimodal interaction, Audiovisual feedback, Sonification, Physical modeling synthesis, Virtual reality, Augmented reality, Human-centered design, Human-computer interaction, HCI, Medical images, Medical image interaction, Surgical navigation, Brain surgery, Brain tumor, Tumor localization.}

\teaser{
  \centering
  \includegraphics[width=\linewidth]{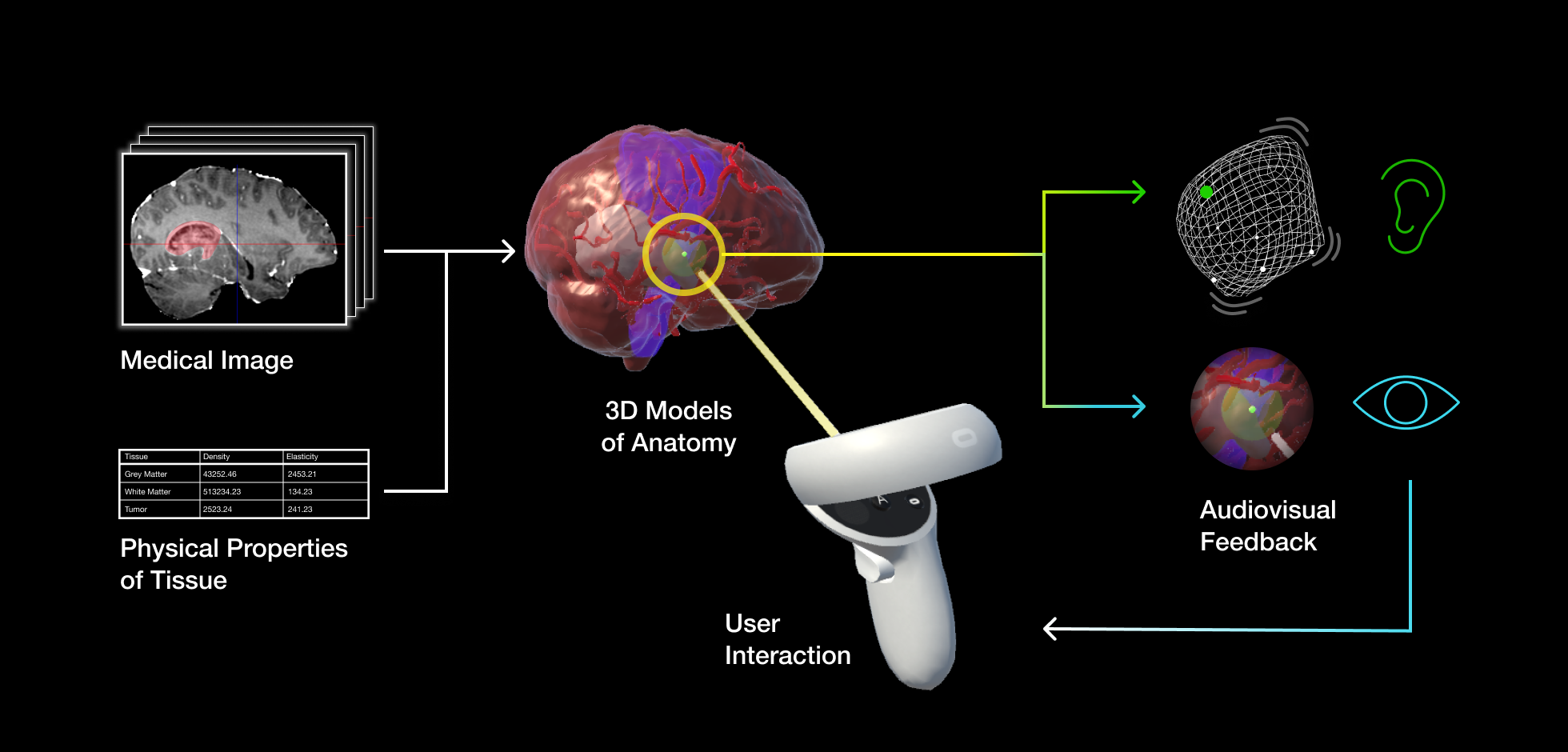}
  \caption{%
  	Physically informed Multimodal Medical Image Interaction (MMII) Framework
  }
  \label{fig:teaser}
}

\graphicspath{{figs/}} 

\usepackage{tabu}                      
\usepackage{booktabs}                  
\usepackage{lipsum}                    
\usepackage{mwe}                       

\usepackage{mathptmx}                  
\usepackage{amsmath}

\begin{document}



\firstsection{Introduction}
\maketitle

Many clinical tasks, ranging from diagnosis to surgery, necessitate sophisticated interaction with multimodal medical image data, presenting a significant challenge due to the intricate nature of these data and the high precision and expertise required. These data are often presented as static volumes and, in some cases, as time-indexed sequences. For instance, a radiologist might be faced with assessing a patient's condition based on Magnetic Resonance Imaging (MRI) scans, a series of two-dimensional (2D) images of the patient's anatomy. Up to three 2D MRI slices need to be navigated and processed simultaneously by the radiologists during diagnosis. Surgical procedures pose an even greater challenge, requiring unparalleled spatial and temporal coordination, precision, and advanced motor skills. This pushes surgeons' cognitive abilities to their limits. Consider the example of brain tumor surgery, where the surgeon must consider medical image data to navigate surgical tools to a tumor via a restricted access point while monitoring numerous critical structures, such as vessels and nerves that closely surround the target area, making the task exceedingly difficult.

A precise and dynamic delivery of crucial information regarding the patient's anatomy and physiology, facilitating robust and reliable perception of high-dimensional medical data, is essential for the success of medical procedures. 
 
Research in cognitive psychology has shown that causally correlated multimodal feedback can activate specific brain regions~\cite{van2008pip,ngo2010auditory}, leading to enhanced task performance and reduced cognitive load~\cite{shams2008benefits}, particularly in tasks involving the localization and tracking of moving objects. The human visual system operates efficiently by organizing objects spatially, yet it struggles with tracking multiple streams of dynamic objects~\cite{dixon2014inattentional}. Furthermore, visual feedback can sometimes distract users from the primary task area and challenge the user's hand-eye coordination if not displayed efficiently. Conversely, the human auditory system excels at perceiving subtle changes in multi-layered data on a fine temporal scale in an omnidirectional manner. We aim to leverage the benefits of the visual and the auditory domain through multimodal interaction. 

While there are works combining audio and visual feedback in Mixed Reality settings for medical tasks\cite{bork2015, joeres2021, schuetz2023audiovis}, they provide simplified interactions through navigational cues on the location or orientation of medical instruments. This interaction approach can be restrictive, as it prescribes specific actions to the user.
Instead, our goal is to augment the medical expert's perception while preserving their decision-making autonomy. To achieve this, we propose a method that enhances their perception of human anatomy during the medical task through multimodal interaction. By introducing an interactive audiovisual technique, we aim to capture the complexity of medical data and intuitively present it to clinicians in a dynamic manner. The proposed modules combine anatomical data derived from biomechanics research with medical imaging data, offering expert users comprehensive and realistic information.

To evaluate the method's effectiveness, we carried out two user studies. The first study aimed to assess the capability of 34 general users to learn correlations between the visual anatomy representation and its auditory cues{, while} the second study focused on determining the usability and accuracy of the method by testing it with nine expert users specialized in neuroradiology and neurosurgery. The results of our studies confirmed that the proposed framework can enhance medical image interaction.\newline

Our work provides the following main contributions:
\begin{enumerate}
    \item We introduce a physically informed multimodal interaction framework for medical experts that provides spatial audiovisual feedback.
    \item We generate auditory representations of anatomical structures using a physically informed sonification model, optimized for real-time applications.
    \item We report results from a user study involving 34 volunteers, {demonstrating} a significant learning effect of the audiovisual representations of anatomical structures.
    \item We report results from a user study with four neurosurgeons and five neuroradiologists proving significantly enhanced accuracy in brain tumor localization when using MMII.
\end{enumerate}

\section{Related Work}

\subsection{Medical Image Representation and Interaction}
Traditionally, {three-dimensional (3D)} medical images, e.g., from computed tomography (CT) imaging or MRI, are displayed as sectional images. Sectional images depict a slice of the body. Compared to 3D anatomy views, they prevent structures from overlapping and thus allow for a clear view of all tissue on the sectional image while improving the examination of soft tissue. Three sectional images are commonly used to navigate the medical image data during diagnosis or treatment tasks. The three standard planes are sagittal, coronal, and axial. The sagittal plane divides the body into the left and right sides. The coronal plane divides the body into posterior (back) and anterior (front) parts, while the axial plane divides the body into its upper and lower parts\cite{madden2008sectionl}.
The downside of cross-sectional anatomy is the lack of three-dimensional spatial relationships. At the same time, interpretation of {3D} structures from anatomical features in cross-sectional images is challenging \cite{ben2022multimodal}.

To overcome these drawbacks, 3D anatomy views have been introduced to enhance the visuospatial perception of anatomical structures. 3D views are advantageous in understanding the position, shape, size, and relationship with neighboring anatomical structures, fostering three-dimensional anatomy understanding. Especially for users with lower visual-spatial abilities, stereoscopic Augmented Reality (AR) 3D views have been shown to improve anatomy learning compared to desktop-based 3D visualizations\cite{bogomolova2020}.
Combining 3D anatomy views and 2D sectional images has also proven to enhance medical image interpretation and anatomy learning\cite{Keenan2020}. In addition, direct manipulation of 3D anatomy views has been reported superior to passively viewing interactions\cite{jang2017}.
Another approach to 3D medical image interaction utilizing embodiment to improve the learning experience is the screen-based AR system called Magic Mirror that enables users to explore anatomy in relation to their own body\cite{blum2012mirracle,bork2017magicmirror, bork2019magicmirror}.
Apart from education, 3D views presented in medical AR \cite{navab2023MedAR} are used in image-guided surgery to enhance surgical planning and decision-making during surgical tasks. Medical AR has proven the potential to increase precision, for example, in orthopedic surgery\cite{andress2018, dennler2020}. {Another study proved significantly enhanced precision when using 3D anatomy views over 2D views for a contouring task during VR-based radiotherapy treatment planning\cite{chen2022}.}

However, most often, the spatial augmentations remain visual. Some works have focused on audiovisual interactions to leverage the potential of multiple senses in conveying spatial information during medical tasks. Among these works is an audiovisual AR system that uses auditory and visuotemporal guidance to improve the 3D localization of occluded anatomy. The study showed enhanced needle placement accuracy when using audiovisual guidance \cite{bork2015}. A work from 2017 showed an audiovisual AR system for laparoscopic procedures. However, no significant improvements of audiovisual over visual AR were reported\cite{joeres2021}. A study that sonified and visualized the position and angle of a magnetic coil used in transcranial magnetic stimulation resulted in improved usability of audiovisual over unimodal, visual guidance \cite{schuetz2023audiovis}. 

Notably, the above studies provided audiovisual feedback on the {placement} of medical instruments rather than information on the nature of the anatomical structures. To the best of our knowledge, no audiovisual feedback based on anatomical structures' geometric and physical properties has been proposed so far.

\subsection{Sonification}

Research has explored non-visual AR solutions~\cite{marquardt2020comparing}, focusing on comparisons between audio-tactile and visual guidance, and has investigated aural augmented reality~\cite{yang2022audio}, highlighting its use in navigation, education, and healthcare. Sonification - the process of conveying data through sound in a systematic, objective, and reproducible way - can significantly improve user experience in interactive systems and interfaces~\cite{hermann2011sonification}. This approach is widely used in interaction design to not only create dynamic and immersive experiences but also to transmit information, offer feedback, and shape user behavior~\cite{franinovic2013sonic}. Sonification techniques such as \textbf{audification} and \textbf{parameter-mapping sonification (PMSon)}, along with more recent developments such as \textbf{model-based sonification (MBS)}~\cite{hermann1999listen,bovermann2006tangible}, have been crucial in transforming diverse data patterns into auditory experiences that are both perceptible and intuitive. These techniques enable users to understand the data, track its changes, gain insights, and detect patterns within it.

In this {work}, we exclusively examine medical applications of interactive sonification, investigating essential methodologies grounded in psychoacoustics as the backbone of these applications.

\textbf{Audification}, the process of directly translating data waveforms into the audible domain, serves as a foundational technique in the medical field, reminding us of traditional tools like the stethoscope or heart rate monitors. This method has been employed to convey simple physiological signals in an auditory form. Researchers have explored audifying bioelectric signals to listen to brain activities~\cite{valjamae2013review}. However, directly converting these data into audio signals often results in noisy and hard-to-understand outputs. Achieving meaningful results through audification typically necessitates advanced, sometimes invasive, sensory technologies capable of extracting significant information from the body at audio rates.

The challenge of generating understandable sounds from data has prompted researchers to turn to alternative sensory modalities, such as visual inputs from cameras, converting them into a simpler, low-dimensional space to establish meaningful mappings to acoustic parameters such as pitch and amplitude. The method, called \textbf{parameter mapping sonification (PMSon)}, leverages an explicit mapping function to accurately control the output, reducing noise and enhancing the clarity of the auditory representation. PMSon has become the preferred technique in sonification, especially in the medical field.

In medical sonification, PMSon has been widely used in translating the position or state of navigated surgical instruments relative to particular structures~\cite{wegner1997surgical,matinfar2017surgical,roodaki2017sonifeye}, converting spatial characteristics of medical imaging data into acoustic features~\cite{ahmad2010sonification}, and in transforming signals derived from surgical devices, such as blood loss monitors or oximeters, into sound~\cite{matinfar2019sonification}. PMSon has proven particularly effective in augmenting image-guided navigation, offering notable improvements in accuracy and guidance~\cite{hansen2013auditory,matinfar2023sonification,ziemer2023three}.

Fundamental studies~\cite{dubus2013systematic} on the impact of mapping strategies in sonification have identified pitch as the primary auditory dimension commonly used in mappings. Efforts to expand mapping from one-dimensional~\cite{parseihian2016comparison} and two-dimensional data\cite{ziemer2017psychoacoustic,ziemer2018psychoacoustical,matinfar2023sonification} to three orthogonal dimensions~\cite{ziemer2019three,ziemer2023three} have shown promise in enhancing spatial awareness. However, the learnability and {intuitiveness} of these techniques, particularly for critical applications such as surgery, remain uncertain.

Conversely, the issue of effectively implementing auditory guidance for multidimensional tasks has been addressed {by Kantan et al.}~\cite{kantan2022sound}, suggesting that simplifying multidimensional tasks into unidimensional tasks could lead to more efficient outcomes. In this approach, users process each dimension and its corresponding sound property sequentially rather than dealing with concurrent guidance presentations. This method has {been} shown to reduce completion times and interruptions, ultimately imposing a lower cognitive load.

However, since most medical applications focus on tool navigation, requiring the modeling of data across Cartesian or polar coordinate systems in multidimensional {space}, neither simultaneous multidimensional mapping nor sequentializing into multiple navigation steps is optimal. This complexity hinders the scalability of these models, especially when tracking multiple objects of interest, leading to reduced intuitiveness and prolonged learning periods. Furthermore, the task of enhancing data dimensionality with PMSon to accurately depict complex physical features such as anatomical shapes and textures presents a significant challenge.

Sch\"utz et al.~\cite{schuetz2024shape} developed a method that, {contrary to the }sonification of a tool's distance to a predefined target, employs {an interactive sonification for shape exploration} to convey anatomical shapes, thus improving the understanding of geometrical configurations through sound. Nevertheless, this approach {only provides sonification of 2D shapes, limiting its suitability} for more complex applications.

Yet, sound uniquely enables the conveyance of multilayered information through the simultaneous presentation of multiple data streams, capitalizing on one of its inherent strengths. Auditory sensations, characterized by qualities such as brightness, roughness, fullness, and sensory pleasantness, as described in~\cite{zwicker2013psychoacoustics}, are inherently multidimensional. This is indeed due to their reliance on the spectral distribution of frequencies. Such multidimensionality greatly expands the scope for developing sonification models and mappings, allowing for a comprehensive representation of data complexity through detailed mappings. 

\textbf{Model-based sonification (MBS)} transforms high-dimensional data into auditory models, facilitating the representation of intricate data patterns, including medical image data. As highlighted in~\cite{matinfar2023tissue}, this method incorporates the multifaceted nature of medical images into a sound-based framework, producing a rich and nuanced auditory output. Despite its potential, the feasibility of applying these techniques in real-time interactive scenarios remains uncertain, calling for further investigation to fully understand and leverage their capabilities in dynamic environments. The construction of sound models in MBS is commonly achieved through physical modeling synthesis, as documented by~\cite{phmss}. This technique aims to mimic the physical characteristics of real-world instruments, effectively emulating the behavior of actual objects. A prominent instance of such mathematical simulations is the Modalys software environment~\cite{eckel1995sound}, which facilitates physical modeling synthesis through the application of the finite element method. This involves solving differential equations related to a vibrating system, allowing the model to capture the system's dynamic attributes, including natural frequencies, damping factors, and mode shapes. 

{This} literature review highlights a gap in systems that provide dynamic, interactive feedback that effectively incorporates the physical characteristics of human anatomy in a distinct, multi-layered, and intuitive manner. There is a lack of systems that are easy to learn, combining both auditory and visual feedback. In discussing sonification for medical applications, it is crucial to consider the physics of the underlying anatomical structures. This consideration is essential for designing models that are not only extendable but also maintain system intuitiveness. Furthermore, the majority of existing studies have adopted a broad approach without sufficiently incorporating the users' background knowledge, particularly their medical expertise.

\begin{table*} [h!]
\caption{Physical properties of the anatomical structures used in Study 1 based on research in biomechanics\protect\footnote{https://itis.swiss/virtual-population/tissue-properties/database/density/}\protect\cite{bartlett2016biomechanical,bartlett2020mechanical,kang2024viscoelastic}: stiffness (Young modulus (kPa)), density (kg/m\textsuperscript{3}), Poisson's ratio}
\centering
\label{tab:physicalprop}
\begin{tabular}{llrrr}
    \toprule
     & Tissue & Young modulus (kPa) & Density (kg/m³) & Poisson's ratio \\
    \midrule
    Rigid & Vertebra (bone) & 50 - 20,000 & 1908 $\pm$ 133 & 0.2 - 0.3 \\
    & Blood Vessel Wall & 1,000 - 20,000 & 1102 $\pm$ 64 & 0.2 - 0.3 \\
    & Intervertebral Disc (cartilage) & 0.3 - 1,000 & 1100 $\pm$ 1 & 0.1 - 0.49 \\
    Semi-rigid & Spinal Cord & 0.5 - 20 & 1075 $\pm$ 52 & 0.45 - 0.49 \\
    Soft  & White Matter & 0.5 - 3.5 & 1041 $\pm$ 2 & 0.4 - 0.5 \\
    & Grey Matter & 0.5 - 2.5 & 1045 $\pm$ 8 & 0.4 - 0.5 \\
    & Glioma (brain tumor) & 0.1 - 1.5 & tumor grade dependent & 0.3 - 0.5 \\
    \bottomrule
    \end{tabular}
\end{table*}

\section{Multimodal Medical Image Interaction Framework}
We propose a multimodal medical image interaction (MMII) framework that provides dynamic audiovisual feedback during interaction with medical image data (Figure \ref{fig:teaser}). Unlike conventional unimodal interaction methods this audiovisual feedback provides detailed information on the anatomy's geometry, texture and size. The MMII framework takes medical images and physical properties of the respective tissue as input to create a visualization and sonification model. Through {user} interaction with these models real-time physically informed audiovisual feedback is generated.

\subsection{Causality-Informed Framework}
Medical professionals have a deep understanding of complex anatomical structures through extensive education and practical experiences. This knowledge enables them to effectively associate auditory representations with anatomical data, provided the sonification model has causal relevance to the context. This aspect highlights the importance of user-centered design, which accounts for users' pre-existing knowledge to enhance their interpretation and interaction with medical data.

We propose a causality-informed method for anatomical sonification, focusing on the energetics of sound events - the diverse energetic flows involved in generating sonic phenomena. Sound, in reality is causal, event-based, and relational. It occurs when energetic phenomena oscillate at specific rates and intensities, with these waves propagating across multiple activity zones. A sound event is considered to have occurred only when listeners introduce their perceptual point of access to the soundscape through their entanglement with its complex development. Our approach challenges the often technical abstraction-focused literature, advocating for a relational ontology that considers the energetic, matter, resonance, and perception-action interplay. Designing
a causality-informed framework requires addressing key ontological questions: What generates sound, why does it occur, and how do listeners engage with it?

The causality-informed method enables us to use proximity-based sonification in an efficient and distinctive manner. Instead of treating proximity as an abstract n-dimensional concept that initiates sound production upon changes in relational proximity, we integrate anatomical knowledge from the interaction context. By localizing and adjusting the perceptual access point to objects based on the location of the user-guided interaction point, we aim to improve the perceptibility of proximity. This approach is considered particularly effective for dynamically active phenomena such as heart rate variability. Following a causal approach, we facilitate user interaction with a resonating model to deduce structural insights from sonic feedback. Thus, distinct timbral differences arise between objects with varying textures and structures, engaging listeners as active participants in the exploration process.
ers as active participants in the sound generation process.

\subsection{Human Anatomy and Tissue Characteristics}

Human anatomy constitutes the input to the MMII framework. Beginning with chemicals and progressing to cellular structures, we observe a hierarchical organization in the human body. At a higher level of abstraction, tissues serve as the foundational elements of all anatomical structures within the human body. The complex human organism comprises four basic tissue types: nervous, epithelial, muscle, and connective tissue. Examples of nervous tissue are the spinal cord or grey matter in the brain. An example of epithelial tissue is the simple squamous epithelium forming the wall of an artery. The cardiac muscle is an exemplary muscle tissue; connective tissue can be found in bones, blood, and even cartilage. These tissues are the building blocks of every organ, such as the spinal cord and the brain, which in turn form organ systems, exemplified by the human central nervous system.

\begin{figure*} [h!]
  \centering
    \includegraphics[width=\textwidth]{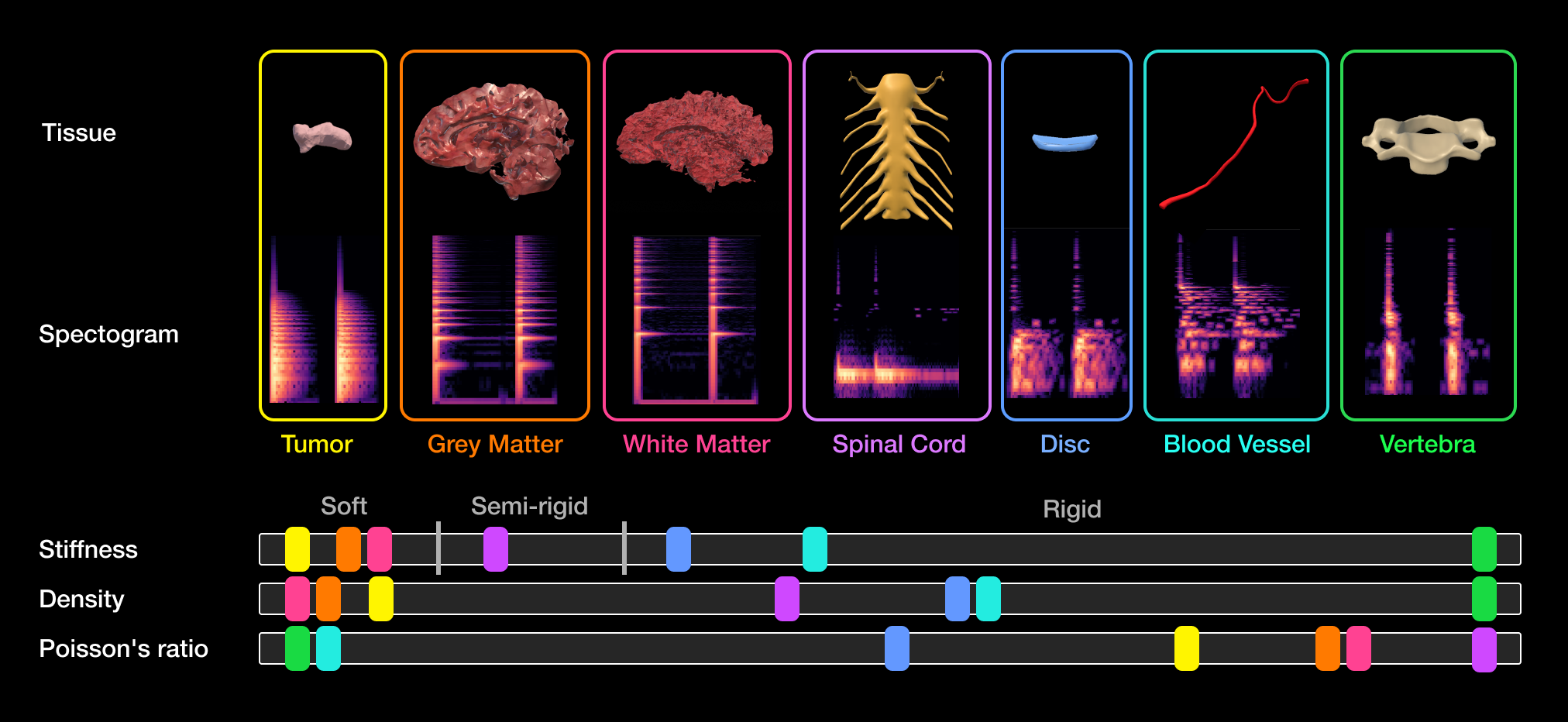}
  \caption{Schema of the tissues used in Study 1 along with their physical properties of stiffness, density and poisson's ratio as well as their resulting mel spectograms.}
  \label{fig:physicalPropsFig}
\end{figure*}

Tissues act as the foundational fabric of the body, intricately woven into the smallest units vital for modeling anatomical structures. These structures significantly affect the dynamic behavior of organisms and are essential in physical modeling sound synthesis. Utilizing the mechanical properties of tissues, such as density, elasticity, energy loss, and Poisson's ratio, enables the simulation and creation of anatomically relevant soundscapes. There is a significant body of research in biomechanics providing detailed data on these physical properties of tissue\footnote[1]{https://itis.swiss/virtual-population/tissue-properties/database/density/}~\cite{bartlett2016biomechanical,bartlett2020mechanical,kang2024viscoelastic}, suitable for our modeling purposes, as indicated in Table~\ref{tab:physicalprop}. Figure \ref{fig:physicalPropsFig} depicts a rough placement of the anatomical structures used in Study 1 along their physical properties.

Moreover, the geometrical configuration and morphological characteristics of anatomical structures are pivotal in determining their dynamic behaviors. These aspects significantly impact the processes of visualization and sonification through wave propagation within structures, emphasizing the dynamic interaction between texture, shape, and function in biological systems. Such interconnections are critical in the development of the proposed causality-informed framework. Medical imaging techniques, including CT, MRI, and ultrasound, allow for visualizing anatomical structures by transforming signals from the imaging system into image intensity values. This transformation facilitates the identification of shapes and patterns, making these methods essential for accurately depicting the geometrical configurations of anatomical structures and providing vital data for their precise mapping.

\subsection{Visualization Model}
Following a causality-informed approach, the visualization model aims to portray a realistic representation of the geometry of human tissue. Segmentations of task-relevant anatomical structures are obtained from the medical image data. The segmentations were turned into surface meshes.
Triggered by user interaction, real-time visual feedback was created. Among dynamic visual changes, we experimented with scaling and colorization effects. Colorization effects were achieved by adjusting the albedo value of the meshes' materials in Unity\footnote{Unity (https://unity.com/)}. Besides changes in material or color, we can imagine a wide array of visual feedback. 3D model deformation could, for example, be suitable for achieving causality-informed visual feedback.

\subsection{Interaction}
A causal relation is also maintained in the user's interaction with the MMII framework by adopting an event-based interaction. Just like playing an instrument creates time-based output from time-based user input, a single input event to the MMII framework will also create a singular, real-time output. Analogous, continuous user input will result in a continuous audiovisual output.

\subsection{Sonification Model}
The proposed sonification model draws inspiration from the general framework introduced in~\cite{matinfar2023tissue}, yet distinguishes itself through the following points:

The framework outlined in~\cite{matinfar2023tissue} presents a broad definition, claiming that any medical image data can be sonified using this approach. The process involves extracting all requisite details for model definition directly from medical images, leading to a highly intricate and detailed modeling process. {It introduces a predominantly conceptual design framework that, while intriguing from a methodological standpoint, provides limited examples to showcase its applicability. Moreover, its feasibility, particularly for real-time applications and rapid prototyping for realistic experimental scenarios, remains unexplored.}

To address these challenges, we streamlined the modeling process by utilizing two distinct datasets: the physical properties of the targeted tissue (Table \ref{tab:physicalprop}) and the geometrical shape of the relevant anatomical structures (Figure \ref{fig:physicalPropsFig}). These structures are derived from medical images, as described in the previous section. They are defined as a polygon mesh, a collection of vertices, edges, and faces that determines the shape of an object using a finite number of quadrilaterals. This setup allows for the numerical solution of the model, particularly the calculation of the displacement of each vertex, through differential equations. Consequently, the model can capture the system's dynamic attributes, including natural frequencies, damping factors, and mode shapes. To initialize the model, the physical properties of the targeted tissue, including elasticity, density, and Poisson's ratio, are mapped to the model parameters. 

Through this, we optimize the model's complexity while ensuring it retains the necessary effectiveness in generating distinct acoustic responses suitable for the proposed application. {As a result, this approach expands the scope of sonification modeling, encompassing a diverse array of anatomical structures for real-time applications.}

For sound generation, the model must be excited by applying force to one or more vertices. This enables the calculation of each node's displacement in an oscillatory form, which is then extracted as an audio signal from one or more vertices. Altering the positions of the input and output vertices significantly affects the resulting sound. Utilizing this model, we create acoustic virtual models capable of producing anatomically informed audio signals through user interaction. This approach integrates crucial anatomical information into the synthesis process, offering users rich and realistic insights. Finally, through minimal real-time preprocessing of the signals, including pitch shifting and amplification, we aimed to enhance the perceptibility of the sounds. For implementing this approach, we utilized the Modalys software~\cite{eckel1995sound}. The audio signals produced through this technique are represented in Figure~\ref{fig:physicalPropsFig}, using a Mel-spectrogram. Readers are also encouraged to refer to the supplementary video to listen to the corresponding audio examples.

There's a notable scenario involving structures that are inherently dynamic, such as blood flow, which naturally generates vibrations without any external interaction. In these cases, we use a signal as the model excitor, specifically generated by a regular heartbeat. Employing advanced sampling techniques such as granular synthesis, we adjust the fine-grained samples of a prerecorded audio file of Korotkoff sounds. This adjustment mimics the variations in blood flow relative to heart rate, tool position, and the surrounding anatomical structures of the blood vessels. Korotkoff sounds are captured using a stethoscope or Doppler device when a blood pressure cuff, positioned distal to the cuff, modifies arterial flow~\cite{porter2001cambridge}. Since the circulation depends on the structural composition of the vasculature, we utilize the granular synthesizer to stimulate a model that mimics the vessel walls. This method allows for the simulation of two distinct scenarios: the surgical tool's proximity to the vessels and the cutting of the vessels. It does so by altering the audio based on whether or not the physical model is engaged.

\section{Experiments}
We conducted two user studies to incrementally test our interaction framework. The first study evaluated the perceptibility of the audiovisual feedback in an isolated manner. To separate the task complexity from the audiovisual feedback, we chose a passive interactive setup in the form of an online study. This enabled us to test the general feasibility of the framework and the suitability of physical modeling {synthesis} to represent human anatomy in the auditory domain. 
In the second study, we were interested in evaluating the potential of the proposed method in a more realistic interaction setup, where the user is provided with multimodal feedback while focusing on a time-sensitive medical task in a Virtual Reality environment. Study 2 focused on the usability and accuracy of the framework for a medical localization task, more precisely, brain tumor localization. The two studies sought to answer the following research questions:

\begin{itemize}
    \item Can users learn the audiovisual correspondence of the auditory and visual representation of anatomy? (Study 1)
    \item Is physical modeling {synthesis} a suitable sonification approach to create distinguishable auditory representations of anatomy? (Study 1)
    \item Can audiovisual interaction improve the usability and accuracy of a medical localization task? (Study 2)
\end{itemize}


\section{Study 1 - Multimodal Correspondence Learning}
We chose to conduct an online study to test the learnability of our framework and determine how well the different anatomical structures can be distinguished based on sound. {The online questionnaire was taken on desktop and laptop computers.} To test our framework for a wide range of human tissue, we chose two areas of the body containing structures of varying physical properties: the spine and the brain. Four diverse spine structures and three brain structures were selected: vertebras (bone), vertebral arteries (blood vessels), intervertebral discs (cartilage), spinal cord (nerves), grey matter, white matter, and brain tumor.

\subsection{Study Implementation}
The online questionnaire was created with SoSci Survey\footnote{SoSci Survey (https://www.soscisurvey.de/)}. Video clips and sound samples were used to simulate the audiovisual interaction. The video and sound files are recordings of a system built in Unity and Max\footnote{Max (https://cycling74.com/products/max)}. 
Two scenes were prepared in Unity, one containing the structures of the cervical spine\footnote{Cervical Spine 3D Model (https://rigmodels.com/)} and one containing the structures of the brain. The 3D brain models were created from MRI image data in the University of California San Francisco Preoperative Diffuse Glioma MRI (UCSF-PDGM) dataset publicly available on the National Cancer Institute's Cancer Imaging Archive\footnote{Cancer Imaging Archive (https://www.cancerimagingarchive.net/)}. FAST, FMRIB's Automated Segmentation Tool\footnote{FAST (https://fsl.fmrib.ox.ac.uk/fsl/fslwiki/FAST)} was used to segment a mesh of the brain's white matter and grey matter from one of the 3D MRI images in the dataset. A 3D object of the brain tumor segmentation was already provided as part of the patient data. Blender\footnote{Blender (https://www.blender.org/)} was used to clean and edit the brain meshes prior to Unity import. The meshes were visualized using surface shaders rendered using physically-based rendering in Unity (Version 2022.3.17f1 LTS). The use of raycasting and mesh colliders enabled a mouse click interaction. To create visual feedback, the objects' size and material changed on click. At the same time UnityOSC\footnote{UnityOSC (https://thomasfredericks.github.io/UnityOSC/)}, a Unity plugin based on Open Sound Control (OSC), a network protocol for interactive computer music \cite{wright2005}, was used to send messages from Unity to Max.
In Max, the incoming OSC messages were routed to the tissue's corresponding physical model to generate the audio feedback. Modalys\footnote{Modalys (https://support.ircam.fr/docs/Modalys/current/)} for Max, a software for virtual musical instruments based on physical models, was used to create instruments from the 3D anatomy models. The Modalys models were parameterized using the physical properties of the different tissues found in Tabel \ref{tab:physicalprop}.

One significant challenge we encountered was that certain structures did not produce audible vibrations or were particularly difficult to sonify due to their inherent characteristics. For instance, brain tissue, being soft tissue, often resulted in sounds that were noisy or inaudible. Even when these sounds are audible, they might not be easily interpretable. To address this, we {modified} the signal into an audible and perceptible sound space by transforming, scaling, and sometimes inverting the information, while crucially preserving the relational context of these structures in comparison to other objects within the scene. This approach ensures that users can perceive the data while maintaining the essential causal relationship of the underlying structures.

\subsection{Participants}
34 volunteers with a mean age of 32.4 $\pm$ 5.9 years participated in the study. Twelve indicated to be women and 22 to be men. The study included participants with medical and non-medical backgrounds. Their occupations ranged from radiologist to biomedical researcher, graphic designer, and music teacher. None of the participants indicated to suffer from a hearing or vision impairment. Regarding the participants' musical experience, 12 indicated listening to music but not making music themselves, 13 indicated playing one instrument, five indicated playing more than one instrument on a regular basis, and four indicated being professional musicians.

\subsection{Study Procedure}
The online study followed a within-subject study design. The participants were first asked some questions about their demographic background, such as age, gender, their medical and musical knowledge{, before they were} guided through a training section explaining the interaction framework and {providing} videos showcasing the simulated audiovisual interaction with all the spine and brain structures. The training section was not time-constrained, and the participants had the chance to get acquainted with the sounds and visuals until they felt confident enough to proceed to the test section. 

\begin{figure*} [h!]
  \centering
    \includegraphics[width=\textwidth]{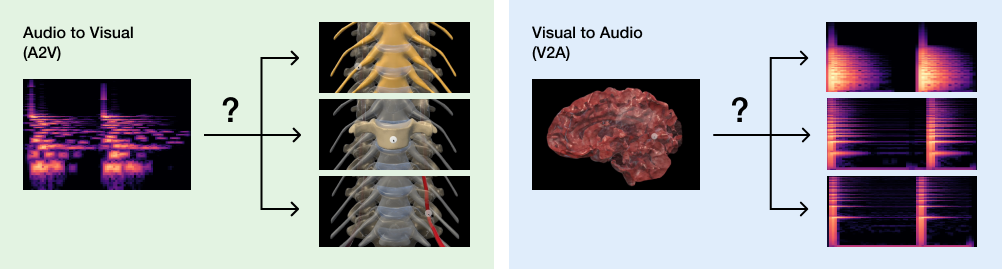}
  \caption{Two question types in Study 1: Audio to Visual (A2V) and Visual to Audio (V2A). In A2V a sound was {played} and its visual correspondence had to be {selected}; in V2A a visual was {shown} and its audio correspondence had to be selected {from a choice of three audio files. Mel spectograms are used to symbolize the audio files in this figure.}}
  \label{fig:study1}
\end{figure*}

The test section was subdivided into two parts: Audio to Visual and Visual to Audio (Figure \ref{fig:study1}). 
In the \textbf{Audio to Visual (A2V)} part the participants were provided with a sound sample and asked to choose the correct visual correspondence from a set of three videos without sound. 
In the \textbf{Visual to Audio (V2A)} part, they were given a video clip without sound, for which they had to select the corresponding audio file from a set of three answer options. Both parts, A2V and V2A, contained seven questions followed by a raw NASA-TLX\cite{hart1988, hart2006}.
{The two question types were contained in both stages of the study: single and multiple structures. The stages were executed in the order mentioned. In the single structure phase, the participants had to select one visual corresponding to one sound (A2V) or vice versa (V2A); in the multiple structures phase, they had to select the correct sequence of two sounds corresponding to the order of two visuals (V2A) or vice versa (A2V).}
The order of {the A2V and V2A parts} was randomized within the single and multiple structures stage. The online questionnaire concluded with a few open questions to capture qualitative feedback on the visual and sound design. 

\subsection{Results}
\subsubsection{Performance and Task Load}
Participants whose rates of correct answers fell outside the range of the mean $\pm$ three standard deviations were considered outliers. Our criteria for outlier detection did not identify any outliers. Given the binary nature of the study's output values, with responses categorized as either 'correct' or 'incorrect,' we employed McNemar's test to assess the statistical significance of the correct answers across two paired conditions: V2A vs. A2V {(Table~\ref{tab:results_study1_nasa_modality})}, and single vs. multiple structures {(Table~\ref{tab:results_study1_nasa_single_multiple})}. 
McNemar's test revealed a statistically significant {effect of modality on response accuracy} ($p<0.01$). {A2V resulted in a significantly higher rate of correct answers than V2A {(Table~\ref{tab:results_study1_nasa_modality})}.} Similarly, a comparison between {the} single and multiple {structures stage} demonstrated a significant effect ($p<0.001$), suggesting that the number of structures influenced response correctness {{(Table~\ref{tab:results_study1_nasa_single_multiple})}}.

\begin{figure}[h!]
  \centering
    \includegraphics[width=0.48\textwidth]{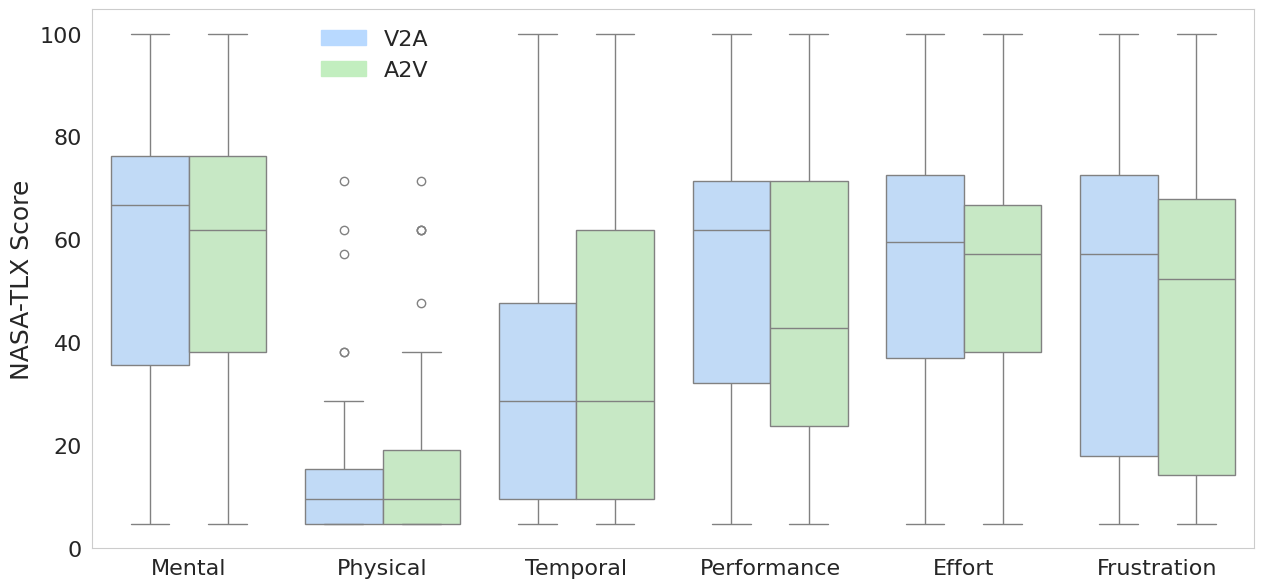}
  \caption{{NASA-TLX task load ratings {for} Visual to Audio (V2A) and Audio to Visual (A2V) question types in Study 1.}}
  \label{fig:study1-nasa-modality}
\end{figure}

\begin{figure}[h!]
  \centering
    \includegraphics[width=\linewidth]{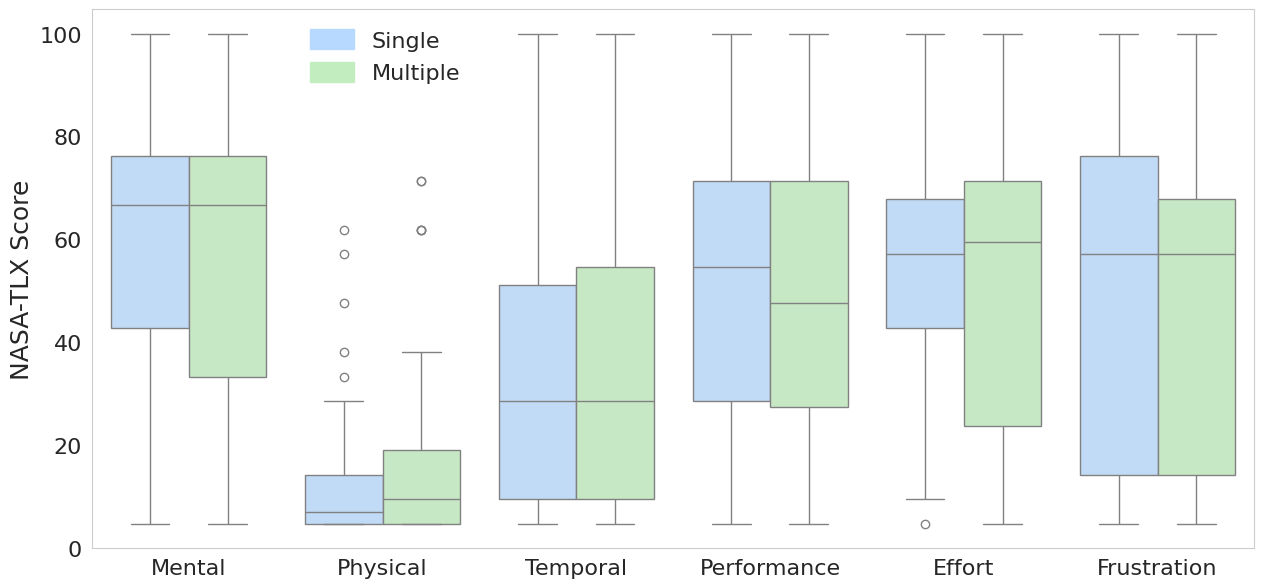}
  \caption{{NASA-TLX task load ratings grouped by {the} single and multiple structures stage in Study 1.}}
  \label{fig:study1-nasa-single-multiple}
\end{figure}

We conducted paired samples t-tests for each NASA-TLX subscale since the assumption of normality for the differences between paired scores was verified using the Shapiro-Wilk test. 
The analysis revealed significant differences ($p<0.001$) in overall task load (NASA-TLX) between V2A and A2V tasks {(Table~\ref{tab:results_study1_nasa_modality})}. Participants reported significantly lower {overall task load} ($p<0.001$) and Frustration ($p<0.001$), {and significantly better} Performance ($p<0.001$) for the A2V task when compared to V2A. Conversely, Mental Demand ($p<0.05$), Physical Demand ($p<0.05$), Temporal Demand ($p<0.05$){, and Effort ($p<0.001$)} were rated significantly higher under the A2V condition {(Figure~\ref{fig:study1-nasa-modality})}. {No significant differences in task load were reported for the single versus multiple structures stage (Figure~\ref{fig:study1-nasa-single-multiple}).}

\begin{table} [b!]
\caption{Means and standard deviations of rate of correct answers (in \%), overall task load and NASA-TLX subscales ([0, 100]; smaller better) for the V2A and A2V {question types} in Study 1 summarized across the single and multiple structures stages }
\label{tab:results_study1_nasa_modality}
\footnotesize
\centering
\begin{tabular}{lrr}
\toprule
n = 34  & Visual to Audio & Audio to Visual\\
\midrule
        Rate of correct answers (\%) & 0.67 $\pm$ 0.47  & 0.75 $\pm$ 0.44\\
        NASA-TLX {- Overall} & 42.76 $\pm$ 30.37 & 41.78 $\pm$ 29.52\\
        NASA-TLX - Mental Demand & {55.64 $\pm$ 28.69} & {57.48 $\pm$ 26.10}\\
        NASA-TLX - Physical Demand & {13.03 $\pm$ 14.19} & {14.23 $\pm$ 15.92}\\
        NASA-TLX - Temporal Demand & {32.53  $\pm$ 26.34} & {34.05 $\pm$ 28.03}\\
        NASA-TLX - Performance & {52.47 $\pm$ 27.17} & {46.75 $\pm$ 27.76}\\
        NASA-TLX - Effort & {55.06 $\pm$ 27.37} & {52.11 $\pm$ 25.02}\\
        NASA-TLX - Frustration & {48.30 $\pm$ 30.96} & {46.06 $\pm$ 30.80}\\
        \bottomrule
    \end{tabular}
\end{table}

\begin{table} [b!]
\caption{Means and standard deviations of rate of correct answers (in \%), overall task load, and NASA-TLX subscales ([0, 100]; smaller better) for the single and multiple structures stages in Study 1 summarized across {V2A and A2V question types.}}
\label{tab:results_study1_nasa_single_multiple}
\footnotesize
\centering
\begin{tabular}{lrr}
\toprule
n = 32  & Single & Multiple\\
\midrule
        Rate of correct answers (\%) & 0.63 $\pm$ 0.48  & 0.78 $\pm$ 0.42\\
        NASA-TLX {- Overall} & 42.91 $\pm$ 29.36  & 41.63 $\pm$ 30.52\\
        NASA-TLX - Mental Demand & {57.98  $\pm$ 25.41} & {55.14 $\pm$ 29.26}\\
        NASA-TLX - Physical Demand & {12.42 $\pm$ 12.79} & {14.84 $\pm$ 17.00 }\\
        NASA-TLX - Temporal Demand & {33.03  $\pm$ 26.39} & {33.55 $\pm$ 28.00}\\
        NASA-TLX - Performance & {51.25 $\pm$ 27.10} & {47.97 $\pm$ 28.03}\\
        NASA-TLX - Effort & {54.92 $\pm$ 23.43} & {51.75 $\pm$ 28.72}\\
        NASA-TLX - Frustration & {47.86 $\pm$ 30.60} & {46.50 $\pm$ 31.18}\\
        \bottomrule
    \end{tabular}
\end{table}

\subsubsection{Qualitative Feedback}
The participants {found identifying sounds of hard structures like bone easier than soft structures such as brain tissue. They especially requested more apparent sound} differences between the brain tissues (white matter, grey matter, tumor). They noted that some sounds were dominant and distinct, making them easy to recognize. However, others were too similar, and thus, distinction was hampered. Participants were more certain of their answers when the questions involved these dominant, distinct sounds. When asked how to improve {the} sonification, they responded, "make the sounds more different from each other" and "make brain sounds softer". Overall, they found the visualizations were appropriate for the purpose.

\subsection{Discussion Study 1}
Our results showed that users are capable of creating audiovisual associations between visual and audio representations of anatomical structures in a relatively short amount of time {regardless of musical expertise}. A significant increase in correct answers ($p<0.001$) from 63.4\% in stage one (single structures) to 77.9\% in stage two (multiple structures) indicates that the MMII framework is learnable. 
Although the complexity of the task increased in stage 2 (multiple structures) the rate of correct answers increased. The study furthermore answered our second research question by proving the suitability of physical modeling {synthesis} for anatomy sonification. 

The NASA-TLX Performance subscale showed that the participants' subjectively perceived performance {aligned with} their objective performance. 
{They were overall more accurate in the A2V task and the multiple structures stage.}

While the online format helped the generalisability of our results by reaching a larger and more diverse group of participants, offering a direct instead of a passive interaction might have improved the transferability and relevance of the Study 1 results to Study 2. {Furthermore, a study comparing musical experts with non-musicians might result in interesting findings.}


\begin{figure*} [h!]
  \centering
    \includegraphics[width=\textwidth]{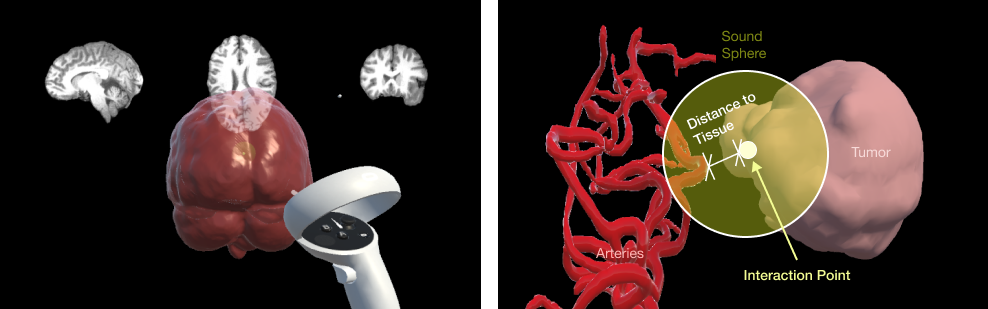}
  \caption{Left: Scene view inside the head-mounted display of the three medical slices and the brain during the visual and audiovisual (MMII) condition in Study 2. Right: Illustration of the sound sphere - the audible range around the point of interaction at the end of the controller ray. The distance from the interaction point to the anatomical structures defines the amplitude of the structures' sound.}
  \label{fig:study2}
\end{figure*}

\section{Study 2 - Multimodal Brain Tumor Localization}
After having proven that MMII has the potential to establish audiovisual associations for anatomy structures, we conducted a second study to examine the framework's capabilities for a real medical task. We evaluated the framework for brain tumor localization with neurosurgeons and neuroradiologists. Localization of the tumor area is, for example, performed during diagnosis or surgery. During diagnosis, the tumor is identified, and its location is determined in a radiology report. During surgery, the tumor needs to be localized to plan a surgical approach that will prevent damage to functional regions of the brain and determine the volume to be resected from the patient's brain. This task requires simultaneous integration of information from several sources, such as slice views, instrument location, and monitoring numerous critical structures in the brain. This task pushes surgeons' cognitive abilities and thus presents a stellar use case for multimodal feedback.
To test the localization task in a simulated manner, we built a Virtual Reality (VR) application for the Meta Quest 2\footnote{Meta Quest 2 (https://www.meta.com/de/quest/products/quest-2/)}. We compared conventional visual medical image interaction with audiovisual interaction using MMII.

\subsection{Study Implementation}
The application was built in Unity and Max. To replicate the standard interaction with medical image data, we included slice views of the three planes (axial, coronal, sagittal) into the VR environment using the UnityVolumeRendering package\footnote{UnityVolumeRendering (https://github.com/mlavik1/UnityVolumeRendering)}(Figure \ref{fig:study2}). In addition, we placed a 3D model of the brain into the Unity scene. The slice views and the brain model were created from 3D MRI images of the UCSF-PDGM dataset. nii2mesh\footnote{nii2mesh (https://github.com/neurolabusc/nii2mesh)} was used to convert the NIfTI 3D voxel images of the tumor and brain segmentations to triangulated meshes.
Besides the grey matter and tumor segmentation, we further included models of the cerebral arteries as well as the right and left corticospinal tract (motor pathways). These critical structures must not be cut during surgery, therefore we decided to include them in our VR environment.

A controller-based interaction with the brain anatomy was chosen for the localization task. {Unity XR Interaction Toolkit (version 2.5.2) was used.} An interaction at a fixed distance from the controller prevented the controller from occluding the brain model visualization. Two concentric spheres were attached to the endpoint of a ray in line with the pointing direction of the controller (Figure \ref{fig:study2}). The inner white sphere marked the current point of interaction with the brain model and determined the location of the three slicing planes. The large green sphere determined the audible radius around the point of interaction. All structures inside the sound sphere were sonified. The amplitude of the sounds was influenced by the distance between the interaction point and the closed point on the object's mesh inside the sound sphere. The distance was normalized for the radius of the sphere. 

The audio feedback was triggered by sending OSC messages from {the Android application (running on a Meta Quest 2)} to Max {(running on a laptop)}. The messages included the names of the structures inside the sound sphere and their respective distance to the point of interaction. Again, Modalys for Max was used to create physical models from the anatomy structures. Unlike the previous study, we also made use of granular synthesis to create a pulsating sound to represent the cerebral arteries.
{All sounds were synthesized in Max and played via stereo speakers connected to the laptop with Aux cable. The streaming latency was minimal and not perceptible to the human ear.}

\subsection{Participants}

Nine medical doctors took part in the study, one woman and eight men. The participants had an average age of 30.1 years, with a standard deviation of 2.8 years. The study included one senior neurosurgeon and three attending neurosurgeons, as well as one senior neuroradiologist and four attending neuroradiologists. While the neurosurgeons indicated performing brain tumor surgery on a daily to weekly basis, the neuroradiologists stated to take brain MRIs and create radiology reports on a daily to weekly basis. None of the doctors indicated to have a vision or hearing impairment. Five participants had never used Augmented Reality (AR) or Virtual Reality (VR) before, two had used AR or VR once before, and two indicated to have used AR or VR a couple of times. Two doctors indicated listening to music but not playing any instrument. Three indicated that they knew how to play an instrument but rarely played it, and three {physicians} answered that they {regularly} played one or multiple instruments.

\subsection{Study Procedure}
In this comparative study that followed a within-subject design, the medical doctors used the conventional visual interaction and the novel audiovisual interaction to localize tumors in a simulated VR setting. The users were first given an introduction to the MMII framework and the study procedure. After completing a short survey assessing demographic background (age, gender, profession), medical and {musical} expertise, the training phase of the first of two study conditions started. During training, the doctors could familiarize themselves with the VR environment, the controller interaction, and the visual and audio feedback. The same task was performed in both conditions. The medical users had to localize and mark the tumor volume by evenly placing spheres on the surface of the tumor. They had to localize six tumors per condition. Both the order of the tumors and the order of the visual and audiovisual conditions were randomized. After each condition, the doctors filled out a raw NASA-TLX and a few specific questions regarding the task and the (audio)visual interaction.

\subsection{Data Analysis}
We used an overlap-based metric to determine the accuracy of the tumor localization task. The overlap of the marked tumor area (MT) with the ground truth tumor area (GT) was measured using the Sørensen–Dice coefficient, the most commonly used metric in validating medical volume segmentations\cite{taha2015}. We stored both the location of the 3D tumor objects and the locations of the user-placed marking spheres in the Unity coordinate frame. We used the sphere locations as input points to create a mesh using the ConvexHull function in SciPy\footnote{SciPy (https://scipy.org/)}. The volume of this hull was used as the marked tumor volume in the Dice calculation.

\begin{equation}
\textrm{Dice\ coefficient} =  \frac{2|MT \cap GT|}{|MT| + |GT|}
\end{equation}

\subsection{Results}
\subsubsection{Performance and Task Load}
\begin{figure}[ht!]
  \centering
    \includegraphics[width=\linewidth]{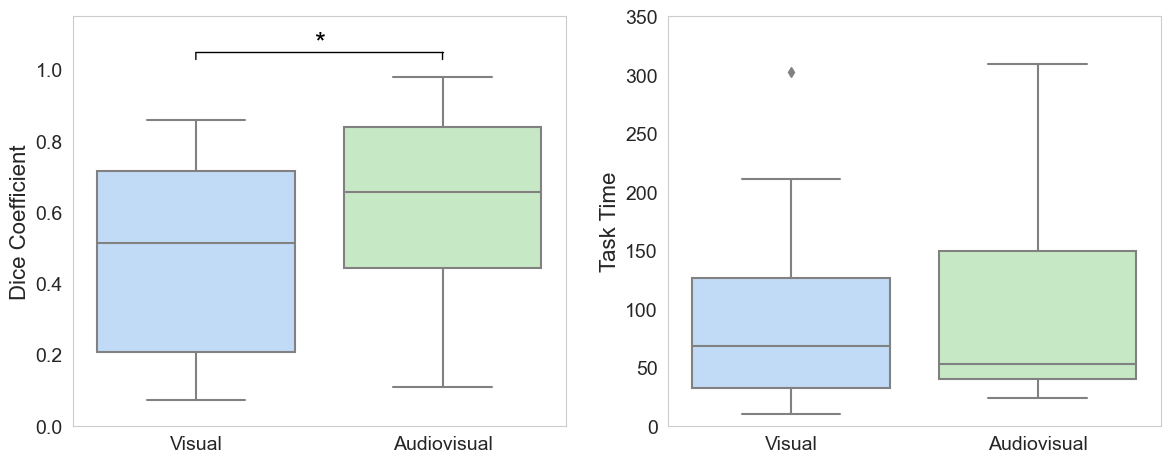}
  \caption{{Box plots of dice coefficient and task time per trial for both conditions in Study 2: Visual and Audiovisual (MMII); * = p < 0.05.}}
  \label{fig:study2-plots}
\end{figure}

A Shapiro-Wilk test showed the normality of the Dice and NASA-TLX data. Values outside the mean $\pm$ three standard deviations were considered outliers {and removed from the sample. To account for differing sample lengths after outlier removal, values were randomly sampled from the larger sample to match the size of the smaller sample.} Since the study followed a within-subjects design, a paired samples t-test was used to test for significance. A significant difference was found between the visual and audiovisual Dice data {(Figure \ref{fig:study2-plots})}. The tumor markings placed using the audiovisual feedback resulted in a significantly higher Dice coefficient value ($p<0.05$), indicating improved localization accuracy when using the audiovisual compared to the visual feedback.

A Shapiro-Wilk test showed a non-normal distribution of the task time data. Eight outliers were removed using the interquartile range method. A Wilcoxon signed rank test showed no significant difference in task time per tumor between the visual and audiovisual condition {(Figure \ref{fig:study2-plots})}.

\begin{table} [h!]
\caption{Means and standard deviations of localization accuracy and task load for the visual and audiovisual condition in Study 2: Sørensen–Dice coefficient ([0,1]; larger better), overall task load, individual NASA-TLX subscales ([0, 100]; smaller better, except Performance - larger better), task time in seconds.}
\label{tab:results_study2}
\footnotesize
\begin{tabular}{lrr}
\toprule
n = 9  & Visual & Audiovisual (MMII) \\
\midrule
S{\o}rensen–Dice Coefficient & {0.48 $\pm$ 0.25} & {0.60 $\pm$ 0.28} \\
        NASA-TLX {- Overall} & 59.07 $\pm$ 14.32 & 52.96 $\pm$ 10.99\\
        NASA-TLX - Mental Demand & {82.22 $\pm$ 9.72} & {68.89 $\pm$ 15.36}\\
        NASA-TLX - Physical Demand & {48.89 $\pm$ 17.64} & {44.44 $\pm$ 13.33}\\
        NASA-TLX - Temporal Demand & {53.34 $\pm$ 27.84} & {48.89 $\pm$ 24.72}\\
        NASA-TLX - Performance & {48.89 $\pm$ 20.89} & {54.44 $\pm$ 13.33}\\
        NASA-TLX - Effort & {68.89 $\pm$ 21.47} & {61.11 $\pm$ 15.37}\\
        NASA-TLX - Frustration & {52.23 $\pm$ 29.91} & {40.00 $\pm$ 18.71}\\
        Task Time & {83.73 $\pm$ 66.24} & {100.40 $\pm$ 83.84}\\
        \bottomrule
    \end{tabular}
\end{table}

No significant difference was found for the {task load (NASA-TLX)} results. Table~\ref{tab:results_study2} lists the means and standard deviations of the Dice, {task load, and task time results} for the visual and audiovisual (MMII) condition.

\subsubsection{Qualitative Feedback}
The participants stated that the sonification helped them to better perceive the distance to the tumor. They further said that the change in amplitude depending on the distance was helpful. When asked about ways to improve the audiovisual interaction, they suggested a "more nuanced soundscape," "more distinct sounds," and "a sharp onset of a different sound" once the interaction point enters the tumor volume to allow for more precise localization. The majority of the participants felt that the sound characterized the different anatomical structures very well. When asked which of the two conditions they would like to use in the future, all nine doctors answered to prefer the audiovisual feedback.

\subsection{Discussion Study 2}
Our study showed significantly increased ($p<0.05$) task accuracy when using MMII. Although the localization accuracy improved when using the audiovisual feedback, the Dice coefficient is still relatively low. This aligns with the doctors' suggestions to adapt the audio feedback to be more sensitive to subtle changes in location. Such adaptation would {be needed to increase the precision of} MMII {and make it} suitable for {application in real} surgical precision tasks.

{While works such as the one by Chen et al.\cite{chen2022} showcase the annotation precision that can be achieved using 3D visualizations for surgical planning, our work elicits the benefit of reduced mental demand and frustration when using audiovisual interaction during real-time surgical interactions. Combining the advantages of both 3D visualizations and audiovisual interaction could guarantee spatial precision while enabling perception of dynamic changes during surgeon-anatomy interactions, leading to increased task performance and confidence.}

{From a balanced distribution of musical experience among the medical participants, we can assume that the framework is accessible regardless of musical expertise.}

We further saw non-significant differences in task load and a non-significant increase in task time when using multimodal over unimodal interaction. A study involving a larger group of medical experts could give insights into the validity of the observed tendency. A further limitation of our study is the unequal representation of gender. A future study should try to achieve a balance of gender among the {medical} participants to ensure that our claims are valid for the entire population of users.


\section{Discussion}
We extensively tested the interaction framework in two user studies. Eight tissues with {varying} physical properties and characteristics, from rigid (vertebra) to soft (brain tissue) {and} from static (vertebra) to dynamic (cerebral arteries), {were sonified}. We {saw} that a physical modeling {synthesis} approach that takes the anatomy's geometry and physical properties into account is suitable to create intuitive and diverse sounds. However, we also experienced that some tissues due to their similarity in shape and physical properties, e.g., grey matter and white matter of the brain create sounds that are hard to distinguish for the general user. In those cases, we propose normalization, scaling, and transformation of the tissues' physical values into an audibly distinguishable range.

Although the presented visual feedback was received well by the participants, we would suggest a more rigorous application of the causality-informed nature of the framework for the visualization model, e.g., by equally dynamic visual responses to changes in anatomy geometry and physical properties.

\subsection{Limitations}
Although we tried to diversify the range of anatomical structures from rigid to soft, we acknowledge that this framework has so far only been tested on structures of the spine and the brain. Evaluation of this method on further parts of the body would be required to claim general application for all human tissues.

Another limitation of our work is the evaluation of MMII for medical tasks in a simulated VR environment. Although useful for initial testing of the method it lacks realism. Evaluating MMII on an AR HMD and providing a physical model of the patient to the medical experts would certainly increase the relevance of the results to clinical practice. 

\subsection{Outlook}
Building upon the findings of this {work}, future {studies should evaluate the framework in an actual clinical setting and include a wider range of anatomical structures. Future work} could {also} focus on incorporating more detailed data from tissue as suggested in~\cite{matinfar2023tissue}. However, this requires optimized methodologies capable of incorporating such detailed information in finite element models, while providing low-latency feedback suitable for surgical applications at a fine temporal scale. Additionally, intraoperative diagnostic techniques such as mass spectroscopy, Raman spectrometry~\cite{van2022diagnostic}, {or acoustic listening~\cite{illanes2018novel,ostler2020acoustic} could provide real-time tissue information as input to the MMII framework and thus further enrich the audiovisual feedback.}

Furthermore, incorporating physiological data, which is inherently dynamic and more complex, has the potential to add significant value to the {MMII framework}. Physiology provides a more holistic view of the body's dynamics. For instance, an event in the heart could have effects at a distal position, influencing tissues elsewhere in the body. In this work, {a blood vessel sound was proposed that demonstrates an inital physiologically-based approach for blood flow sonification. Offering} dynamic feedback on data such as cerebral blood flow measured through MRI or metabolic changes observed through PET scans {would present additional applications} that showcase the advantages of the proposed multimodal feedback method.

\section{Conclusion}

The Multimodal Medical Image Interaction (MMII) framework showcased in this {work} presents a promising method for engaging with human anatomy through audiovisual means. The proposed physical modeling approach to anatomy sonification has proven to be easy to learn and to characterize anatomical structures well. 
Our work presented an exemplary use case of physically informed multimodal interaction in a surgical task. Evaluation involving nine medical {doctors} indicated increased accuracy in brain tumor localization when using MMII.

Our findings underscore the advantages of audiovisual interactions compared to unimodal, visual feedback, highlighting the potential of multimodal approaches as viable alternatives to traditional medical image interaction methods. We hope this work will inspire more investigation and broader adoption of multimodal {interactions} in medical applications and beyond.



\bibliographystyle{abbrv-doi-hyperref}

\bibliography{template}

\begin{thebibliography}{10}

\bibitem{ahmad2010sonification}
A.~Ahmad, S.~G. Adie, M.~Wang, and S.~A. Boppart.
\newblock Sonification of optical coherence tomography data and images.
\newblock {\em Optics Express}, 18(10):9934--9944, 2010.

\bibitem{andress2018}
S.~Andress, A.~Johnson, M.~Unberath, A.~F. Winkler, K.~Yu, J.~Fotouhi, S.~Weidert, G.~Osgood, and N.~Navab.
\newblock On-the-fly augmented reality for orthopedic surgery using a multimodal fiducial.
\newblock {\em Journal of Medical Imaging}, 5(2):021209--021209, 2018.

\bibitem{bartlett2016biomechanical}
R.~D. Bartlett, D.~Choi, and J.~B. Phillips.
\newblock Biomechanical properties of the spinal cord: implications for tissue engineering and clinical translation.
\newblock {\em Regenerative medicine}, 11(7):659--673, 2016.

\bibitem{bartlett2020mechanical}
R.~D. Bartlett, D.~Eleftheriadou, R.~Evans, D.~Choi, and J.~B. Phillips.
\newblock Mechanical properties of the spinal cord and brain: Comparison with clinical-grade biomaterials for tissue engineering and regenerative medicine.
\newblock {\em Biomaterials}, 258:120303, 2020.

\bibitem{ben2022multimodal}
A.~Ben~Awadh, J.~Clark, G.~Clowry, and I.~D. Keenan.
\newblock Multimodal three-dimensional visualization enhances novice learner interpretation of basic cross-sectional anatomy.
\newblock {\em Anatomical sciences education}, 15(1):127--142, 2022.

\bibitem{blum2012mirracle}
T.~Blum, V.~Kleeberger, C.~Bichlmeier, and N.~Navab.
\newblock mirracle: An augmented reality magic mirror system for anatomy education.
\newblock In {\em 2012 IEEE Virtual Reality Workshops (VRW)}, pp. 115--116. IEEE, 2012.

\bibitem{bogomolova2020}
K.~Bogomolova, I.~J. van~der Ham, M.~E. Dankbaar, W.~W. van~den Broek, S.~E. Hovius, J.~A. van~der Hage, and B.~P. Hierck.
\newblock The effect of stereoscopic augmented reality visualization on learning anatomy and the modifying effect of visual-spatial abilities: A double-center randomized controlled trial.
\newblock {\em Anatomical sciences education}, 13(5):558--567, 2020.

\bibitem{bork2017magicmirror}
F.~Bork, R.~Barmaki, U.~Eck, P.~Fallavolita, B.~Fuerst, and N.~Navab.
\newblock Exploring non-reversing magic mirrors for screen-based augmented reality systems.
\newblock In {\em 2017 IEEE virtual reality (VR)}, pp. 373--374. IEEE, 2017.

\bibitem{bork2015}
F.~Bork, B.~Fuers, A.-K. Schneider, F.~Pinto, C.~Graumann, and N.~Navab.
\newblock Auditory and visio-temporal distance coding for 3-dimensional perception in medical augmented reality.
\newblock In {\em 2015 IEEE International Symposium on Mixed and Augmented Reality}, pp. 7--12. IEEE, New York, NY, USA, 2015. \href{https://doi.org/10.1109/ISMAR.2015.16}
{doi: {{%
10\hspace{.1pt}\discretionary{.}{%
}{.}\hspace{.4pt}1109\discretionary{/}{%
}{/}ISMAR\hspace{.1pt}\discretionary{.}{%
}{.}\hspace{.4pt}2015\hspace{.1pt}\discretionary{.}{%
}{.}\hspace{.4pt}16}}}


\bibitem{bork2019magicmirror}
F.~Bork, L.~Stratmann, S.~Enssle, U.~Eck, N.~Navab, J.~Waschke, and D.~Kugelmann.
\newblock The benefits of an augmented reality magic mirror system for integrated radiology teaching in gross anatomy.
\newblock {\em Anatomical sciences education}, 12(6):585--598, 2019.

\bibitem{bovermann2006tangible}
T.~Bovermann, T.~Hermann, and H.~Ritter.
\newblock Tangible data scanning sonification model.
\newblock In {\em Proceedings of the 12th International Conference on Auditory Display}, 2006.

\bibitem{chen2022}
C.~Chen, M.~Yarmand, V.~Singh, M.~V. Sherer, J.~D. Murphy, Y.~Zhang, and N.~Weibel.
\newblock Vrcontour: Bringing contour delineations of medical structures into virtual reality.
\newblock In {\em 2022 IEEE International Symposium on Mixed and Augmented Reality (ISMAR)}, pp. 64--73, 2022. \href{https://doi.org/10.1109/ISMAR55827.2022.00020}
{doi: {{%
10\hspace{.1pt}\discretionary{.}{%
}{.}\hspace{.4pt}1109\discretionary{/}{%
}{/}ISMAR55827\hspace{.1pt}\discretionary{.}{%
}{.}\hspace{.4pt}2022\hspace{.1pt}\discretionary{.}{%
}{.}\hspace{.4pt}00020}}}


\bibitem{phmss}
P.~R. Cook.
\newblock Physically informed sonic modeling (phism): Percussive synthesis.
\newblock In {\em Proceedings of the 1996 International Computer Music Conference}, pp. 228--231. The International Computer Music Association, 1996.

\bibitem{dennler2020}
C.~Dennler, L.~Jaberg, J.~Spirig, C.~Agten, T.~G{\"o}tschi, P.~F{\"u}rnstahl, and M.~Farshad.
\newblock Augmented reality-based navigation increases precision of pedicle screw insertion.
\newblock {\em Journal of orthopaedic surgery and research}, 15:1--8, 2020.

\bibitem{dixon2014inattentional}
B.~J. Dixon, M.~J. Daly, H.~H. Chan, A.~Vescan, I.~J. Witterick, and J.~C. Irish.
\newblock Inattentional blindness increased with augmented reality surgical navigation.
\newblock {\em American journal of rhinology \& allergy}, 28(5):433--437, 2014.

\bibitem{dubus2013systematic}
G.~Dubus and R.~Bresin.
\newblock A systematic review of mapping strategies for the sonification of physical quantities.
\newblock {\em PloS one}, 8(12):e82491, 2013.

\bibitem{eckel1995sound}
G.~Eckel.
\newblock Sound synthesis by physical modelling with modalys.
\newblock {\em Proc. ISMA'95}, pp. 478--482, 1995.

\bibitem{franinovic2013sonic}
K.~Franinovic and S.~Serafin.
\newblock {\em Sonic interaction design}.
\newblock Mit Press, 2013.

\bibitem{hansen2013auditory}
C.~Hansen, D.~Black, C.~Lange, F.~Rieber, W.~Lamad{\'e}, M.~Donati, K.~J. Oldhafer, and H.~K. Hahn.
\newblock Auditory support for resection guidance in navigated liver surgery.
\newblock {\em The International Journal of Medical Robotics and Computer Assisted Surgery}, 9(1):36--43, 2013.

\bibitem{hart2006}
S.~G. Hart.
\newblock Nasa-task load index (nasa-tlx); 20 years later.
\newblock In {\em Proceedings of the human factors and ergonomics society annual meeting}, vol.~50, pp. 904--908. Sage publications Sage CA: Los Angeles, CA, 2006.

\bibitem{hart1988}
S.~G. Hart and L.~E. Staveland.
\newblock Development of nasa-tlx (task load index): Results of empirical and theoretical research.
\newblock In {\em Advances in Psychology}, vol.~52, pp. 139--183. Elsevier, Amsterdam, Netherlands, 1988.

\bibitem{hermann2011sonification}
T.~Hermann, A.~Hunt, J.~G. Neuhoff, et~al.
\newblock {\em The sonification handbook}, vol.~1.
\newblock Logos Verlag Berlin, 2011.

\bibitem{hermann1999listen}
T.~Hermann and H.~Ritter.
\newblock Listen to your data: Model-based sonification for data analysis.
\newblock {\em Advances in intelligent computing and multimedia systems}, 8:189--194, 1999.

\bibitem{illanes2018novel}
A.~Illanes, A.~Boese, I.~Maldonado, A.~Pashazadeh, A.~Schaufler, N.~Navab, and M.~Friebe.
\newblock Novel clinical device tracking and tissue event characterization using proximally placed audio signal acquisition and processing.
\newblock {\em Scientific reports}, 8(1):12070, 2018.

\bibitem{jang2017}
S.~Jang, J.~M. Vitale, R.~W. Jyung, and J.~B. Black.
\newblock Direct manipulation is better than passive viewing for learning anatomy in a three-dimensional virtual reality environment.
\newblock {\em Computers \& Education}, 106:150--165, 2017.

\bibitem{joeres2021}
F.~Joeres, D.~Black, S.~Razavizadeh, and C.~Hansen.
\newblock Audiovisual {AR} concepts for laparoscopic subsurface structure navigation.
\newblock In {\em Graphics Interface 2021}, 2021.

\bibitem{kang2024viscoelastic}
W.~Kang, L.~Wang, and Y.~Fan.
\newblock Viscoelastic response of gray matter and white matter brain tissues under creep and relaxation.
\newblock {\em Journal of Biomechanics}, 162:111888, 2024.

\bibitem{kantan2022sound}
P.~R. Kantan, S.~Dahl, and E.~G. Spaich.
\newblock Sound-guided 2-d navigation: Effects of information concurrency and coordinate system.
\newblock In {\em Nordic Human-Computer Interaction Conference}, pp. 1--11, 2022.

\bibitem{Keenan2020}
I.~D. Keenan and M.~Powell.
\newblock {\em Interdimensional Travel: Visualisation of 3D-2D Transitions in Anatomy Learning}, pp. 103--116.
\newblock Springer International Publishing, Cham, 2020.

\bibitem{madden2008sectionl}
M.~E. Madden.
\newblock {\em Introduction to sectional anatomy}.
\newblock Lippincott Williams \& Wilkins, 2008.

\bibitem{marquardt2020comparing}
A.~Marquardt, C.~Trepkowski, T.~D. Eibich, J.~Maiero, E.~Kruijff, and J.~Sch{\"o}ning.
\newblock Comparing non-visual and visual guidance methods for narrow field of view augmented reality displays.
\newblock {\em IEEE Transactions on Visualization and Computer Graphics}, 26(12):3389--3401, 2020.

\bibitem{matinfar2019sonification}
S.~Matinfar, T.~Hermann, M.~Seibold, P.~F{\"u}rnstahl, M.~Farshad, and N.~Navab.
\newblock Sonification for process monitoring in highly sensitive surgical tasks.
\newblock In {\em Proceedings of the Nordic Sound and Music Computing Conference 2019 (Nordic SMC 2019)}, 2019.

\bibitem{matinfar2017surgical}
S.~Matinfar, M.~A. Nasseri, U.~Eck, H.~Roodaki, N.~Navab, C.~P. Lohmann, M.~Maier, and N.~Navab.
\newblock Surgical soundtracks: Towards automatic musical augmentation of surgical procedures.
\newblock In {\em Medical Image Computing and Computer-Assisted Intervention- MICCAI 2017: 20th International Conference, Quebec City, QC, Canada, September 11-13, 2017, Proceedings, Part II 20}, pp. 673--681. Springer, 2017.

\bibitem{matinfar2023tissue}
S.~Matinfar, M.~Salehi, S.~Dehghani, and N.~Navab.
\newblock From tissue to sound: Model-based sonification of medical imaging.
\newblock In {\em International Conference on Medical Image Computing and Computer-Assisted Intervention}, pp. 207--216. Springer, 2023.

\bibitem{matinfar2023sonification}
S.~Matinfar, M.~Salehi, D.~Suter, M.~Seibold, S.~Dehghani, N.~Navab, F.~Wanivenhaus, P.~F{\"u}rnstahl, M.~Farshad, and N.~Navab.
\newblock Sonification as a reliable alternative to conventional visual surgical navigation.
\newblock {\em Scientific Reports}, 13(1):5930, 2023.

\bibitem{navab2023MedAR}
N.~Navab, A.~Martin-Gomez, M.~Seibold, M.~Sommersperger, T.~Song, A.~Winkler, K.~Yu, and U.~Eck.
\newblock Medical augmented reality: Definition, principle components, domain modeling, and design-development-validation process.
\newblock {\em Journal of Imaging}, 9(1), 2023.

\bibitem{ngo2010auditory}
M.~K. Ngo and C.~Spence.
\newblock Auditory, tactile, and multisensory cues facilitate search for dynamic visual stimuli.
\newblock {\em Attention, Perception, \& Psychophysics}, 72(6):1654--1665, 2010.

\bibitem{ostler2020acoustic}
D.~Ostler, M.~Seibold, J.~Fuchtmann, N.~Samm, H.~Feussner, D.~Wilhelm, and N.~Navab.
\newblock Acoustic signal analysis of instrument--tissue interaction for minimally invasive interventions.
\newblock {\em International Journal of Computer Assisted Radiology and Surgery}, 15:771--779, 2020.

\bibitem{parseihian2016comparison}
G.~Parseihian, C.~Gondre, M.~Aramaki, S.~Ystad, and R.~Kronland-Martinet.
\newblock Comparison and evaluation of sonification strategies for guidance tasks.
\newblock {\em IEEE Transactions on Multimedia}, 18(4):674--686, 2016.

\bibitem{porter2001cambridge}
R.~Porter.
\newblock {\em The Cambridge illustrated history of medicine}.
\newblock Cambridge University Press, 2001.

\bibitem{roodaki2017sonifeye}
H.~Roodaki, N.~Navab, A.~Eslami, C.~Stapleton, and N.~Navab.
\newblock Sonifeye: Sonification of visual information using physical modeling sound synthesis.
\newblock {\em IEEE transactions on Visualization and Computer Graphics}, 23(11):2366--2371, 2017.

\bibitem{schuetz2024shape}
L.~Sch\"{u}tz, T.~El~Chemaly, E.~Weber, A.~T. Doan, J.~Tsai, C.~Leuze, B.~Daniel, and N.~Navab.
\newblock Interactive shape sonification for tumor localization in breast cancer surgery.
\newblock In {\em Proceedings of the CHI Conference on Human Factors in Computing Systems}, CHI '24. Association for Computing Machinery, New York, NY, USA, 2024. \href{https://doi.org/10.1145/3613904.3642257}
{doi: {{%
10\hspace{.1pt}\discretionary{.}{%
}{.}\hspace{.4pt}1145\discretionary{/}{%
}{/}3613904\hspace{.1pt}\discretionary{.}{%
}{.}\hspace{.4pt}3642257}}}


\bibitem{schuetz2023audiovis}
L.~Sch{\"u}tz, E.~Weber, W.~Niu, B.~Daniel, J.~McNab, N.~Navab, and C.~Leuze.
\newblock Audiovisual augmentation for coil positioning in transcranial magnetic stimulation.
\newblock {\em Computer Methods in Biomechanics and Biomedical Engineering: Imaging \& Visualization}, 11(4):1158--1165, 2023. \href{https://doi.org/10.1080/21681163.2022.2154277}
{doi: {{%
10\hspace{.1pt}\discretionary{.}{%
}{.}\hspace{.4pt}1080\discretionary{/}{%
}{/}21681163\hspace{.1pt}\discretionary{.}{%
}{.}\hspace{.4pt}2022\hspace{.1pt}\discretionary{.}{%
}{.}\hspace{.4pt}2154277}}}


\bibitem{shams2008benefits}
L.~Shams and A.~R. Seitz.
\newblock Benefits of multisensory learning.
\newblock {\em Trends in cognitive sciences}, 12(11):411--417, 2008.

\bibitem{taha2015}
A.~A. Taha and A.~Hanbury.
\newblock Metrics for evaluating 3d medical image segmentation: analysis, selection, and tool.
\newblock {\em BMC medical imaging}, 15:1--28, 2015.

\bibitem{valjamae2013review}
A.~V{\"a}ljam{\"a}e, T.~Steffert, S.~Holland, X.~Marimon, R.~Benitez, S.~Mealla, A.~Oliveira, and S.~Jord{\`a}.
\newblock A review of real-time eeg sonification research.
\newblock In {\em International Conference on Auditory Display 2013 (ICAD 2013)}, 2013.

\bibitem{van2008pip}
E.~Van~der Burg, C.~N. Olivers, A.~W. Bronkhorst, and J.~Theeuwes.
\newblock Pip and pop: nonspatial auditory signals improve spatial visual search.
\newblock {\em Journal of Experimental Psychology: Human Perception and Performance}, 34(5):1053, 2008.

\bibitem{van2022diagnostic}
L.~Van~Hese, S.~De~Vleeschouwer, T.~Theys, S.~Rex, R.~M. Heeren, and E.~Cuypers.
\newblock The diagnostic accuracy of intraoperative differentiation and delineation techniques in brain tumours.
\newblock {\em Discover Oncology}, 13(1):123, 2022.

\bibitem{wegner1997surgical}
C.~M. Wegner and D.~B. Karron.
\newblock Surgical navigation using audio feedback.
\newblock In {\em Medicine Meets Virtual Reality}, pp. 450--458. IOS Press, 1997.

\bibitem{wright2005}
M.~Wright.
\newblock Open sound control: an enabling technology for musical networking.
\newblock {\em Organised Sound}, 10(3):193--200, 2005.

\bibitem{yang2022audio}
J.~Yang, A.~Barde, and M.~Billinghurst.
\newblock Audio augmented reality: a systematic review of technologies, applications, and future research directions.
\newblock {\em journal of the audio engineering society}, 70(10):788--809, 2022.

\bibitem{ziemer2023three}
T.~Ziemer.
\newblock Three-dimensional sonification as a surgical guidance tool.
\newblock {\em Journal on Multimodal User Interfaces}, 17:253--262, 2023.

\bibitem{ziemer2017psychoacoustic}
T.~Ziemer, D.~Black, and H.~Schultheis.
\newblock Psychoacoustic sonification design for navigation in surgical interventions.
\newblock In {\em Proceedings of Meetings on Acoustics}, vol.~30. AIP Publishing, 2017.

\bibitem{ziemer2019three}
T.~Ziemer and H.~Schultheis.
\newblock Three orthogonal dimensions for psychoacoustic sonification.
\newblock {\em arXiv preprint arXiv:1912.00766}, 2019.

\bibitem{ziemer2018psychoacoustical}
T.~Ziemer, H.~Schultheis, D.~Black, and R.~Kikinis.
\newblock Psychoacoustical interactive sonification for short range navigation.
\newblock {\em Acta Acustica united with Acustica}, 104(6):1075--1093, 2018.

\bibitem{zwicker2013psychoacoustics}
E.~Zwicker and H.~Fastl.
\newblock {\em Psychoacoustics: Facts and models}, vol.~22.
\newblock Springer Science \& Business Media, 2013.

\end{thebibliography}

\appendix 

\end{document}